\documentclass[10pt, conference]{IEEEtran}

\usepackage{caption}
\usepackage{cite}
\usepackage{url}
\usepackage{amssymb}
\usepackage{blindtext}
\usepackage{comment}
\usepackage{algorithm}
\usepackage{algorithmic}
\usepackage{subcaption}
\usepackage{xcolor,soul,framed}
\usepackage{amsmath}
\usepackage{graphicx}
\usepackage{multirow}
\usepackage{balance}
\usepackage[hidelinks]{hyperref}

\IEEEoverridecommandlockouts
\def\BibTeX{{\rm B\kern-.05em{\sc i\kern-.025em b}\kern-.08em
    T\kern-.1667em\lower.7ex\hbox{E}\kern-.125emX}}

\ifCLASSINFOpdf
\else
\fi

\begin{document}
%
\title{Cascaded Transformer for Robust and Scalable \\SLA Decomposition via Amortized Optimization}

\author{\IEEEauthorblockN{Cyril Shih-Huan Hsu}
\IEEEauthorblockA{Informatics Institute\\
University of Amsterdam\\
Amsterdam, The Netherlands\\
s.h.hsu@uva.nl}
}


%


\maketitle

\begin{abstract}
The evolution toward 6G networks increasingly relies on network slicing to provide tailored, End-to-End (E2E) logical networks over shared physical infrastructures. A critical challenge is effectively decomposing E2E Service Level Agreements (SLAs) into domain-specific SLAs, which current solutions handle through computationally intensive, iterative optimization processes that incur substantial latency and complexity. 
To address this, we introduce Casformer, a cascaded Transformer architecture designed for fast, optimization-free SLA decomposition. Casformer leverages historical domain feedback encoded through domain-specific Transformer encoders in its first layer, and integrates cross-domain dependencies using a Transformer-based aggregator in its second layer. The model is trained under a learning paradigm inspired by Domain-Informed Neural Networks (DINNs), incorporating risk-informed modeling and amortized optimization to learn a stable, forward-only SLA decomposition policy.
Extensive evaluations demonstrate that Casformer achieves improved SLA decomposition quality against state-of-the-art optimization-based frameworks, while exhibiting enhanced scalability and robustness under volatile and noisy network conditions. In addition, its forward-only design reduces runtime complexity and simplifies deployment and maintenance.
These insights reveal the potential of combining amortized optimization with Transformer-based sequence modeling to advance network automation, providing a scalable and efficient solution suitable for real-time SLA management in advanced 5G-and-beyond network environments.

\end{abstract}


%
\IEEEpeerreviewmaketitle

\section{Introduction}
The rollout of 5G and the evolution toward 6G have redefined mobile network infrastructure as a flexible, multi-service platform catering to diverse vertical industries. A central enabler of this transformation is \emph{network slicing}, which allows the instantiation of multiple End-to-End (E2E) logical networks over a shared physical substrate, each governed by a tailored Service Level Agreement (SLA). These SLAs specify performance commitments through Service Level Objectives (SLOs), encompassing key metrics such as latency, throughput, reliability, and security.
A single network slice may span multiple administrative and technological domains, such as the access, transport, and core domain of the network. To ensure consistent delivery of the E2E SLA across these heterogeneous domains, the SLA must be decomposed into domain-specific SLAs that each domain can support. This decomposition is critical to support optimal resource allocation and maintain SLA compliance over the slice lifecycle.
While several works have addressed related challenges, they vary in focus. For example, the IETF draft~\cite{ietf-teas-5g-network-slice-application-03} explores the difficulty of mapping E2E slice requirements to underlying transport infrastructure. In~\cite{hcltech2023networkslicing}, the emphasis is on automation, orchestration, and SLA monitoring, whereas~\cite{iovanna2022networkslicing} discusses the interplay between SLA parameters and transport resource provisioning. Other studies, such as~\cite{su2019resource}, stress the importance of SLA decomposition in resource management, and~\cite{10011552} proposes AI-assisted frameworks to automate this process in future 6G environments. A recent work ~\cite{10515206} emphasizes the need for scalable SLA decomposition across multi-domain and multi-technology environments.

A common architectural approach to managing network slices is through a hierarchical two-level management system, as illustrated in Fig.~\ref{fig:system}. At the top level, a service orchestrator handles the E2E lifecycle of the network slice, while lower-level domain controllers manage the local instantiation of the slice within each domain. The orchestrator is in charge of decomposing the SLA, but it often operates with limited real-time visibility into domain-level states and relies instead on historical performance data. To address this gap,~\cite{Vleeschauwer21_SLAdecomposition} proposes a two-step approach employing non-parametric risk models to estimate the behavior of underlying domains, followed by an optimization-based decomposition. This direction is extended in~\cite{SLADNN23}, which introduces neural network-based risk modeling, and further refined by~\cite{hsu2024online,rails}, which propose dynamic and adaptive frameworks to reflect changing conditions at different levels of granularity. Despite these advances, current approaches still rely on iterative optimization and frequent online model updates, which can incur significant computational overhead and latency, particularly in dynamic and large-scale environments. Moreover, they require maintaining domain-specific risk models throughout operation, making deployment and maintenance more complex. In this work, we propose \emph{Casformer}, a cascaded Transformer architecture that addresses these limitations by amortizing the decomposition process into a single forward pass.
Trained to simulate the optimal decisions produced by existing risk model-based frameworks, Casformer enables a significantly faster SLA decomposition without the need for online optimization or model updates, thereby enhancing scalability and efficiency without sacrificing decision quality, and simplifying deployment.
The main contributions of this paper are:
\begin{itemize}
    \item \textbf{Optimization-Free SLA Decomposition.} We propose Casformer, a novel framework that bypasses iterative optimization by directly predicting SLA decomposition decisions via amortized optimization. By embedding domain knowledge and adopting a structured learning paradigm inspired by Domain-Informed Neural Networks, Casformer delivers efficient and scalable decompositions without compromising high decision quality.
    \item \textbf{Innovative Application of Transformers.} Casformer integrates risk modeling directly into a cascaded Transformer structure. It encodes historical domain-level feedback and dynamically aggregates this information to produce accurate SLA decompositions. This innovative use of Transformers leverages their powerful sequence modeling capabilities for network optimization tasks.
    \item \textbf{Extensive Empirical Validation.} Comprehensive experiments show significant improvements of our approach in inference efficiency, scalability, and generalization compared to state-of-the-art methods, providing strong practical support of amortized optimization and Transformer-based architectures in network optimization contexts.
\end{itemize}
This paper is organized as follows: Section~\ref{sec:related} discusses related work. Section~\ref{sec:system_model} defines the system model and the problem. Section~\ref{sec:method} introduces our proposed framework. Section~\ref{sec:perf_eval} details the simulation setup, and Section~\ref{sec:results} presents and discusses the results. Finally, Section~\ref{sec:conclusion} concludes the paper.

\section{Related Work}\label{sec:related}
\subsection{SLA Management}
Early work on SLA decomposition primarily focused on single-domain environments. The authors in~\cite{4273097} addressed SLA management in cloud systems by translating high-level SLOs into low-level system thresholds, using performance bounds to determine resource needs. In~\cite{8418115}, the authors proposed an SLA management architecture using neural networks to estimate the resource needed to fulfill SLAs in a single-domain network slicing context. On the other hand, the challenges of ensuring SLA guarantees in multi-domain, multi-technology environments have been increasingly recognized. The authors in~\cite{9305243} emphasized the risks of SLA violations in critical applications like healthcare, pointing to the importance of robust SLA assurance across distributed infrastructures.
In~\cite{10515206}, the authors presents an SLA management and orchestration architecture for multi-domain, multi-technology networks, supporting various distributed schemes.
Other studies~\cite{8417711, 8931583} have proposed predictive approaches to assist in SLA management. For instance,~\cite{8417711} introduced a mapping layer to manage radio resources in a localized service area. In~\cite{8931583}, a deep learning-based framework was developed to estimate resource needs under SLA constraints. However, these methods primarily focus on predictive methods or architectural design and do not explicitly address the decomposition of E2E SLAs across multiple domains.


\subsection{E2E SLA Decomposition} To address the challenge of E2E SLA decomposition, several works have explored optimization and learning-based techniques. In~\cite{su2019resource}, the authors studied resource allocation for E2E network slices using mathematical models. More recently,~\cite{9165317} proposed a supervised learning approach that decomposes E2E SLAs into domain-specific SLOs across multiple domains.
A pioneering work in this direction is presented in~\cite{Vleeschauwer21_SLAdecomposition}, where a two-step method is proposed: first using non-parametric risk models to estimate domain behavior based on historical observations, and then applying an optimization procedure to determine the decomposition. This approach is extended in~\cite{SLADNN23}, which replaces the non-parametric risk models with monotonic neural networks to improve efficiency. More recently, online learning-based frameworks have been introduced in~\cite{hsu2024online, rails, cheng2025odin} to better accommodate dynamic and time-varying network environments.

While these approaches demonstrate increasing sophistication and adaptability, they still depend heavily on online optimization and repeated model updates, which can be computationally intensive and sensitive to hyper-parameters. In contrast, our proposed method eliminates the need for real-time optimization by learning a forward-only neural approximation of the decomposition process, offering improved scalability, responsiveness, and generalizability.



\section{System Model}\label{sec:system_model}



\subsection{Problem Description}
\label{sec:problem_description}
SLA decomposition in multi-domain, multi-technology networks has been addressed through several orchestration schemes~\cite{10515206}. In centralized approaches, a cross-domain orchestrator collects global state information and makes decomposition and resource allocation decisions centrally. Semi-distributed schemes retain centralized SLA decomposition but delegate intra-domain resource management to domain controllers. These approaches, however, suffer from scalability limitations, privacy concerns, and single points of failure.
\begin{figure}[h]
\vspace{-4mm}
     \centering
     \includegraphics[width=1\columnwidth, trim=3cm 0.5cm 6.5cm 1cm, clip]{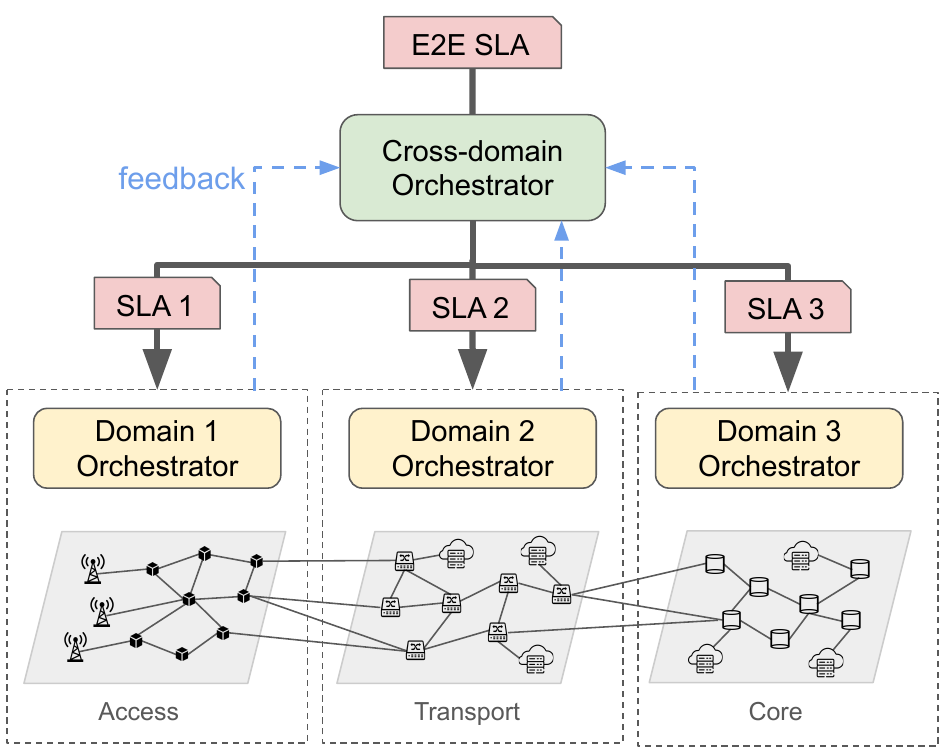}
    \caption{Fully distributed multi-domain network management architecture with feedback-driven decomposition across access, transport, and core domains.}
    \label{fig:system}
    \vspace{-0.5em}
\end{figure}
In contrast, the fully distributed scheme assumes that the cross-domain orchestrator has no visibility into domain-specific states. Instead, it assigns partial SLAs based on historical feedback, while each domain orchestrator independently manages resource allocation. Fig.~\ref{fig:system} illustrates an example of the fully distributed scheme, where the cross-domain orchestrator receives an E2E SLA and decomposes it into partial SLAs for each underlying domain. Domain-specific orchestrators (e.g., Access, Transport, Core) independently manage resource allocation and execution based on their assigned SLAs. Feedback is sent from domains back to the cross-domain orchestrator to inform future decomposition decisions.
This decentralized scheme is more scalable and resilient, but introduces new challenges in ensuring that local decisions collectively satisfy the E2E SLA. Previous work~\cite{rails} has shown that finding the optimal decomposition across domains is an NP-hard problem.
In this study, we focus on this fully distributed setting and aim to design an efficient, adaptive, and optimization-free decomposition framework suitable for real-time operation under domain autonomy.

\subsection{Problem Formulation}
\label{sec:problem_formulation}
\noindent \textbf{SLA Decomposition.} Consider an E2E SLA denoted by a vector $\mathbf{d}_{\text{e2e}}$, which specifies performance bounds on key service metrics such as delay, throughput, and reliability. This E2E SLA must be decomposed into $N$ domains, each operated independently. The decomposition process yields a set of partial SLAs $\{ \mathbf{d}_n \}_{n=1}^{N}$, where $\mathbf{d}_n$ is the SLA portion assigned to domain $n$.

The decomposition must satisfy a set of aggregation rules that ensure the global SLA is preserved. These rules reflect the structure of the service metrics. For instance, the E2E delay $\tau_{\text{e2e}}$ is additive, throughput $\theta_{\text{e2e}}$ is determined by the bottleneck, and reliability $\pi_{\text{e2e}}$ is multiplicative:
\begin{equation}\label{eq:tau}
    \tau_{\text{e2e}} = \sum_{n=1}^{N} \tau_n, \quad \theta_{\text{e2e}} = \min_{n} \theta_n, \quad \pi_{\text{e2e}} = \prod_{n=1}^{N} \pi_n.
\end{equation}
In scenarios where the SLA includes the above metrics, the domain-specific SLA assignment can be expressed as a vector:
\begin{equation}
\mathbf{d}_n = (\tau_n, \theta_n, \pi_n).
\end{equation}

Let $p_n(\mathbf{d}_n)$ denote the acceptance probability of domain $n$ for the SLA portion $\mathbf{d}_n$. This probability is governed by the local domain controller based on internal resource availability, policies, and traffic conditions. Since domains operate independently, we assume the acceptance decisions are statistically independent.
As a result, the E2E acceptance probability of a decomposition $\{ \mathbf{d}_n \}_{n=1}^{N}$ is given by:
\begin{equation}
    p_{\text{e2e}} = \prod_{n=1}^{N} p_n(\mathbf{d}_n).
\end{equation}

The goal of SLA decomposition is to find the set $\{ \mathbf{d}_n \}_{n=1}^{N}$ that satisfies the aggregation constraints while maximizing the probability that all domains accept their assigned SLA portions. Therefore, the decomposition problem can be formulated as:
\begin{equation}\label{eq:decompose}
    \begin{aligned}
        \max_{\{ \mathbf{d}_n \}} \quad & \prod_{n=1}^{N} p_n(\mathbf{d}_n) \\
        \text{s.t.} \quad & \mathbf{d}_{\text{e2e}} = G(\{ \mathbf{d}_n \}), \quad \forall n \in \{1, \dots, N\},
    \end{aligned}
\end{equation}
where $G(\cdot)$ denotes the aggregation function that combines the partial SLAs into the full E2E SLA, as described in~\eqref{eq:tau}. In this study, we focus exclusively on the delay component $\tau$ of the SLA described in~\eqref{eq:tau}. This choice is motivated by both practical and analytical considerations. Delay is one of the most critical performance metrics in E2E service delivery, directly affecting Quality of Service (QoS).
However, the proposed framework in Section~\ref{sec:method} can be straightforwardly extended to handle multi-dimensional SLAs by incorporating appropriate aggregation rules for additional metrics.

Furthermore, in dynamic environments, domain conditions evolve over time due to changes in network state, traffic load, and operational policies, making the acceptance probability $p_n(\cdot)$ time-dependent. We denote this as $p_n^t(\cdot)$ at time step $t$. Accordingly, the SLA decomposition problem in~\eqref{eq:decompose} extends across a time horizon of $T$ steps and is formulated as:
\begin{equation}\label{eq:decompose-time}
    \begin{aligned}
        \max_{\{ \mathbf{d}_n^t \}} \quad & \frac{1}{T} \sum_{t=1}^{T} \prod_{n=1}^{N} p_n^t(\mathbf{d}_n^t) \\
        \text{s.t.} \quad & \mathbf{d}_{\text{e2e}}^t = G(\{ \mathbf{d}_n^t \}), \\
                          & \forall t \in \{1, \dots, T\},\quad \forall n \in \{1, \dots, N\}.
    \end{aligned}
\end{equation}

\vspace{1mm}
\noindent \textbf{Risk Modeling.} In the fully distributed setting, the acceptance of a partial SLA $\mathbf{d}_n$ by domain $n$ is inherently uncertain, as the cross-domain orchestrator lacks visibility into each domain's internal state. Consequently, the acceptance probability $p_n(\cdot)$ is unknown, making it infeasible to directly solve the decomposition problem in~\eqref{eq:decompose}. However, this challenge can be addressed by introducing a domain-specific \emph{risk model} $f_n(\cdot)$ that estimates $p_n(\cdot)$.
In practice, the true acceptance probability $p_n(\cdot)$ is not directly observable and must be inferred from historical feedback. Consider that each domain maintains a dataset of $K$ past observations $\{ (\mathbf{x}_i, y_i) \}_{i=1}^{K}$, where $\mathbf{x}_i$ denotes a previously proposed SLA vector and $y_i \in \{0, 1\}$ indicates the domain's response ($y_i=1$ for acceptance and $y_i=0$ for rejection). A risk model $f_n: \mathbb{R}^d \to [0,1]$ is trained to approximate $p_n(\cdot)$ by minimizing the cross-entropy loss over the dataset:
\begin{equation}\label{eq:bce}
    \mathcal{L}_n = -\sum_{i=1}^{K} \left[ y_i \log f_n(\mathbf{x}_i) + (1 - y_i) \log \left(1 - f_n(\mathbf{x}_i)\right) \right].
\end{equation}

The risk model $f_n$ can take various functional forms depending on the design choice and available data. 
Neural networks are particularly well-suited for capturing subtle dependencies and interactions among different SLA attributes due to their strong capacity for modeling complex relationships between SLA parameters and domain acceptance behavior~\cite{SLADNN23}.

\section{Methodology}
\label{sec:method}

\subsection{Background}\label{sec:method-bg}
\noindent \textbf{RADE.} Real-time Adaptive DEcomposition (RADE)~\cite{hsu2024online} is a state-of-the-art SLA decomposition framework designed to operate in dynamic, fully distributed network environments. Unlike static decomposition methods that rely on fixed risk models, RADE incorporates an online learning mechanism to continuously adapt its decision-making process in response to real-time feedback from domain controllers.
RADE follows a two-step decomposition architecture, building upon prior work~\cite{Vleeschauwer21_SLAdecomposition, SLADNN23}:
\begin{enumerate}
    \item [1.] \textbf{Risk Model Construction.} The cross-domain orchestrator constructs and maintains a domain-specific risk model for each domain, typically implemented as a neural network. These models are trained using historical feedback data, as described in~\eqref{eq:bce}.
    \item [2.] \textbf{SLA Decomposition.} Given an incoming E2E SLA request, the cross-domain orchestrator performs the decomposition by solving the optimization problem defined in~\eqref{eq:decompose}, which aims to maximize the expected E2E acceptance probability using the learned risk models.
\end{enumerate}
The first step leverages gradient-based optimization techniques widely used in deep learning to train the neural network models. The second step combines coarse-grained heuristic search with fine-grained refinement via Sequential Least Squares Programming (SLSQP), enabling the orchestrator to effectively explore the space of feasible decompositions and converge toward a near-optimal solution.

\begin{algorithm}[h]
\caption{RADE Framework}
\label{alg:rade}
\begin{algorithmic}[1]
\REQUIRE Risk models $\{\mathcal{F}_n\}_{n=1}^N$; Memory buffers $\{\mathcal{B}_n\}_{n=1}^N$
\FOR{each time step $t = 1$ to $T$}
    \STATE Collect recent feedback and update $\{ \mathcal{B}_n^t \}_{n=1}^N$
    \STATE \textbf{Update} $\{\mathcal{F}_n\}_{n=1}^N$ via OGD:
    \[
        \{\mathcal{F}_n\}_{n=1}^N \gets \text{OGD}(\{\mathcal{F}_n\}_{n=1}^N, \{ \mathcal{B}_n^t \}_{n=1}^N)
    \]
    \STATE Receive E2E SLA request $\tau_{\text{e2e}}^t$
    \STATE \textbf{Solve} decomposition: 
    \[
        \{ \tau_n^{*,t} \}_{n=1}^N \gets \text{Optimize}(\{\mathcal{F}_n\}_{n=1}^N, \tau_{\text{e2e}}^t)
    \]
    \STATE \textbf{Return} SLA assignments $\{ \tau_n^{*,t} \}_{n=1}^N$
\ENDFOR
\end{algorithmic}
\end{algorithm}


Moreover, to remain effective under time-varying network conditions, RADE continuously updates its risk models via Online Gradient Descent (OGD), using observations collected from recent SLA feedback. To stabilize the learning process and avoid overfitting to short-term fluctuations or corrupted feedback, RADE employs a First-In First-Out (FIFO) memory buffer. This buffer maintains a sliding window of recent feedback, ensuring that the models are updated using a representative and temporally relevant dataset. As a result, RADE is able to maintain more accurate and adaptive risk models over time, which leads to improved decision-making and a higher long-term average E2E acceptance probability, as expressed in~\eqref{eq:decompose-time}. Alg.~\ref{alg:rade} presents the high-level complete steps of the RADE framework. The effectiveness of RADE has been demonstrated in~\cite{hsu2024online}, showing superior performance and robustness against dynamic network environments.



\vspace{1mm}
\noindent \textbf{Transformers.}
Transformers are a class of deep learning models that have revolutionized sequence modeling, particularly in the field of NLP. First introduced in~\cite{vaswani2017attention}, the Transformer architecture replaced traditional Recurrent Neural Networks (RNNs) and Convolutional Neural Networks (CNNs) in many NLP tasks due to its superior scalability, parallelism, and ability to model long-range dependencies. Their ability to handle structured input and long-term dependencies makes them attractive for tasks such as SLA decomposition, where sequences of domain feedback or SLA metrics must be processed holistically.

\vspace{1mm}
\noindent \textbf{Domain‑Informed Neural Network (DINNs).}
DINNs are a class of neural architectures that incorporate structured prior knowledge, physical constraints, or system-specific insights into the learning process~\cite{10.1145/3514228, MORGAN2023938, beucler2021enforcing, quarteroni2025combining}. Unlike purely data-driven models, DINNs leverage domain expertise to guide the learning to improve generalization, where labeled data are limited or expensive to obtain.
The integration of domain knowledge can take several forms, such as:
\begin{itemize}
    \item \textbf{Architectural design:} Embedding domain-specific structures, such as symmetries or conservation laws, into the network layers or connections.
    \item \textbf{Feature encoding:} Preprocessing or embedding raw inputs using known physical, semantic, or geometric relationships to provide more informative representations.
    \item \textbf{Loss regularization:} Introducing physics-informed or logic-based constraints as auxiliary loss terms to ensure outputs comply with known rules.
    \item \textbf{Hybrid models:} Combining data-driven models with analytical or mechanistic models (e.g., simulation engines or differential equations).
\end{itemize}

A prominent example of DINNs is the class of Physics-Informed Neural Networks (PINNs)~\cite{RAISSI2019686}, which incorporate governing physical laws, typically expressed as partial differential equations, directly into the training process. 
This approach has demonstrated strong generalization in regimes where labeled data is scarce but domain knowledge is well understood, highlighting the effectiveness of integrating structured priors into deep learning models. In short, by baking domain insights into the model, DINNs strike a balance between the flexibility of deep learning and the robustness of expert-driven design. 

\vspace{1mm}
\subsection{Risk-Informed Cascaded Transformer}
To address the limitations of online optimization–based two-step SLA decomposition, particularly its sensitivity to noisy feedback, frequent model updates, and complex runtime dependencies, we propose the Risk-Informed Cascaded Transformer (Casformer). Casformer directly predicts SLA decompositions from raw domain-specific feedback in a single forward pass, replacing explicit per-instance optimization with a learned, stable decomposition policy.
Casformer follows a learning paradigm inspired by DINNs, where domain knowledge guides training rather than relying solely on generic supervision. In our implementation, we use RADE, an advanced risk-aware optimization framework, as a \textit{teacher} to generate high-quality decomposition targets. RADE is employed as a representative example due to its robustness and strong performance, but the proposed framework \textbf{is not tied to RADE} and can be trained with any method capable of providing reliable decomposition decisions as supervision. By amortizing the decomposition process into a neural model, Casformer internalizes the structural logic and risk-aware behavior of the \textit{teacher}. Once trained, Casformer operates independently of the \textit{teacher}, simplifying deployment by avoiding continuous optimization and model maintenance.

\begin{figure*}[ht]
    \centering
    \includegraphics[width=0.97\linewidth, trim=0.0cm 0.0cm 0.0cm 0.0cm, clip]{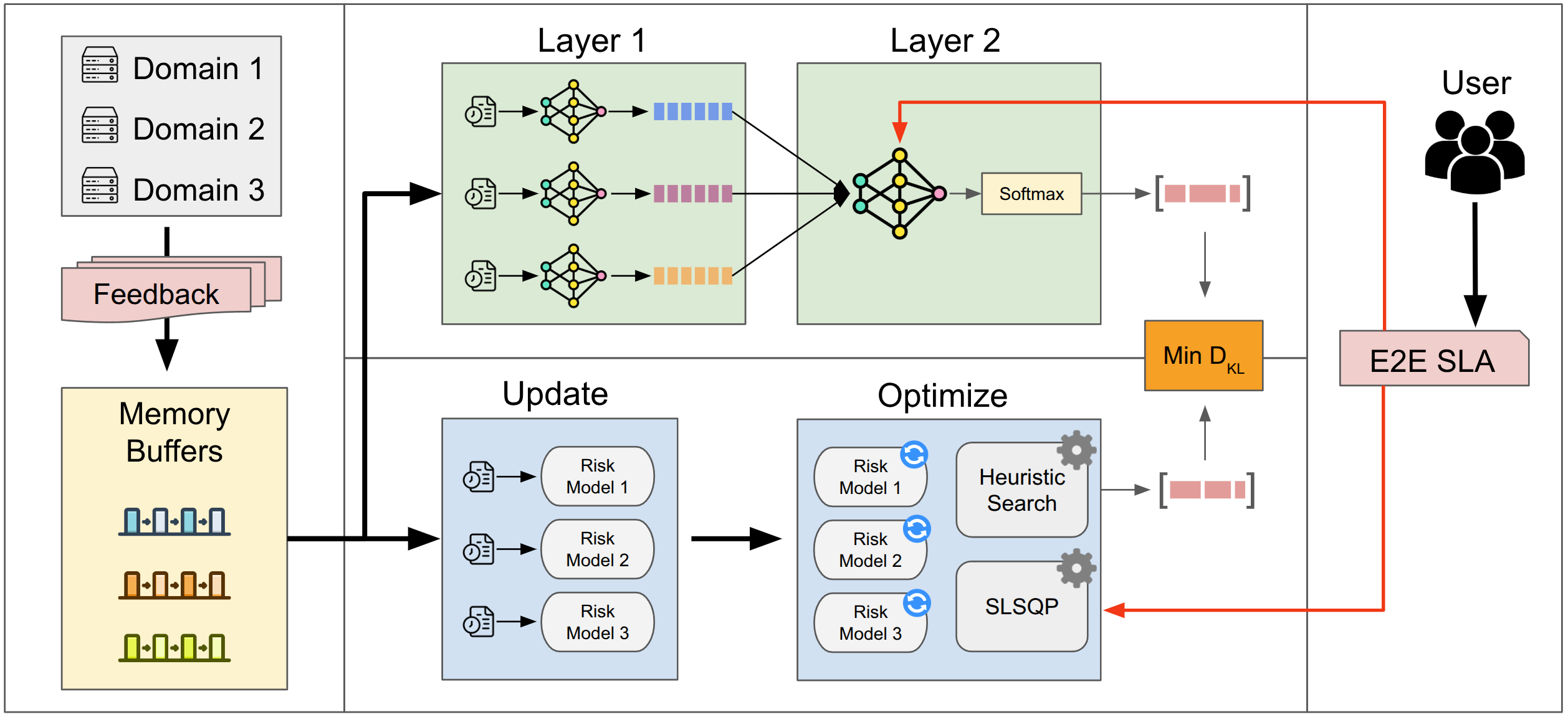}
    \caption{Overview of the Casformer training pipeline guided by RADE supervision. Feedback collected from underlying domains are stored in memory buffers and used to train Casformer (top) by mirroring the decomposition decisions made by RADE (bottom). Casformer learns to predict decomposition ratios via a single forward pass, enabling fast, optimization-free inference.}\label{fig:framework}
    \vspace{-2mm}
\end{figure*}

Fig.~\ref{fig:framework} illustrates the overall architecture of the proposed framework. In particular, on the left, feedback from each domain is continually collected and stored in domain-specific FIFO memory buffers, which retain recent interaction histories. In the bottom-middle section, the RADE module serves as the \textit{teacher} by first updating its domain-specific risk models using the most recent feedback samples. These updated models are then used to solve the SLA decomposition problem through a two-stage optimization procedure combining heuristic search and SLSQP, yielding the optimal per-domain delay assignments. 
At the same time, Casformer acts as the \emph{student} in the top-middle section, receiving the E2E SLA request and processing domain buffers through its first-layer encoders to produce fixed-length representations. These are combined with the E2E SLA budget in the cross-domain aggregator, which eventually outputs decomposition ratios. A KL divergence loss aligns these predictions with the RADE-generated targets, treating RADE as the \emph{teacher}. Both models are triggered simultaneously by the same E2E SLA request, with RADE producing the supervision signal.

\vspace{1mm}
\noindent \textbf{Architecture.} Casformer is structured as a two-layer cascaded transformer designed to encode per-domain SLA feedback and produce a decomposition prediction that aligns with the E2E SLA objective. It consists of two main components: ($i$) domain-specific encoders and ($ii$) a cross-domain aggregator.
\begin{itemize}
    \item \textbf{Memory Buffer.} Similar to RADE, Casformer maintains a memory buffer $\mathcal{B}$ for each domain to store recent SLA interaction records. This buffer operates as a FIFO queue and provides the temporal context required for learning domain behavior over time. Each record in the buffer is a tuple $(x, y) \in \mathcal{B}$ representing a SLA request and its binary acceptance outcome. Importantly, the buffer has limited capacity, and its size plays a critical role in the performance of the model. If the buffer is too large, it may retain outdated data, which can mislead the model and degrade predictive accuracy. Conversely, if it is too small, the model may not have enough context for stable training, leading to volatile predictions. These stored samples form the input sequence for the domain-specific encoders and enable both online and offline training.
    \item \textbf{Domain-Specific feedback Encoders (Layer 1).} In the first layer, each domain $n \in \{1, \dots, N\}$ is assigned a dedicated Transformer encoder $\mathcal{F}^{\text{enc}}_n$ responsible for processing its local feedback history. Each encoder takes as input a set of recent SLA interaction records from the FIFO buffer $\mathcal{B}_n$. The encoder outputs a fixed-length embedding $\mathbf{h}_n \in \mathbb{R}^d$ that summarizes the recent behavior of the corresponding domain: 
    \begin{equation}\label{eq:enc}
    \mathbf{h}_n = \mathcal{F}^{\text{enc}}_n\left(\mathcal{B}_n\right),
    \end{equation}
    where $\mathcal{F}^{\text{enc}}_n(\cdot)$ denotes the encoder associated with domain $n$. This process adopts the architecture of the Universal Sentence Encoder (USE)~\cite{cer2018universal}, which is designed to encode variable-length sequences into fixed-size representations. In our context, the USE-style architecture enables Casformer to flexibly handle different window sizes and domain histories while standardizing their representation for downstream processing.
    \item \textbf{Cross-Domain Aggregator (Layer 2).} In the second layer, a single Transformer encoder acts as a cross-domain aggregator $\mathcal{F}^{\text{agg}}$. It takes as input the set of domain embeddings $\{\mathbf{h}_n\}_{n=1}^{N}$ produced by the first layer as well as the incoming E2E delay SLA $\tau_{\text{e2e}}$, models the dependencies across domains, and generates the final predicted delay decomposition assignment $\{\hat{\tau}_n\}_{n=1}^{N}$.
    Specifically, to incorporate the incoming E2E delay SLA $\tau_{\text{e2e}}$, we first project it into the embedding space using a learnable weight matrix $W_\tau \in \mathbb{R}^{1 \times d}$. The resulting embedding is then added to each domain representation, similar to how positional encodings are applied in standard Transformers:
    \begin{equation}\label{eq:delay_embedding}
        \{\tilde{\mathbf{h}}_n\}_{n=1}^{N} = \{\mathbf{h}_n + W_\tau \cdot \tau_{\text{e2e}}\}_{n=1}^{N}.
    \end{equation}
    Moreover, to enforce the known additive constraint of delay SLAs, i.e., that the partial delays must sum up to the E2E delay as described in~\eqref{eq:tau}, we bake this domain knowledge directly into the model architecture. Rather than predicting raw value of partial delays, the network outputs a softmax distribution over the $N$ domains to represent delay allocation ratios:
    \begin{equation}\label{eq:ratio}
        \hat{\boldsymbol{\phi}} = \text{softmax}\left( \mathcal{F}^{\text{agg}} (\{ \tilde{\mathbf{h}}_n\}_{n=1}^{N}) \right),
    \end{equation}
    where $\hat{\boldsymbol{\phi}} = \{\hat{\phi}_n\}_{n=1}^{N}$ denotes the predicted decomposition ratios. The final per-domain delay assignments are then obtained by scaling these ratios with the given E2E delay SLA:
    \begin{equation}\label{eq:scale}
        \hat{\boldsymbol{\tau}} = \hat{\boldsymbol{\phi}} \cdot \tau_{\text{e2e}},
    \end{equation}
    where $\hat{\boldsymbol{\tau}} = \{\hat{\tau}_n\}_{n=1}^{N}$ denotes the actual predicted decomposition assignment.
    This approach eliminates the need for constraint-penalizing loss terms and ensures that all predictions inherently satisfy the SLA structure.
\end{itemize}

Overall, Casformer can be expressed as a composition of domain-specific encoders, a cross-domain aggregator, and a softmax operation: $\mathcal{F} = \text{softmax} \circ \mathcal{F}^{\text{agg}} \circ \mathcal{F}^{\text{enc}}$.
The attention mechanism in domain-specific encoders enables the model to assign different importance weights to individual feedback samples, which allows Casformer to focus on the most relevant historical information when summarizing domain behavior.
Similarly, the attention mechanism within the cross-domain aggregator allows for dynamically adjusting a partial SLA assignment based on the current conditions of all other domains.
This design also naturally supports variable-length inputs, meaning that the model can handle a changing number of domains without requiring architectural modifications.
Overall, the two-layer design enables Casformer to first extract localized insights per domain and then integrate them holistically to make decomposition decisions. 

\vspace{1mm}
\noindent \textbf{Training Phase.} During training, Casformer is executed in parallel with RADE. For each E2E SLA request that arrives at time step $t$, the following steps are performed:
\begin{enumerate}
    \item [1.] The memory buffer for each domain $n$ provides a set of recent feedback samples.
    \item [2.] Each domain-specific encoder processes its buffer independently, using a transformer-based architecture to encode the feedback sequence into a fixed-length representation as defined in~\eqref{eq:enc}.
    \item [3.] The cross-domain aggregator collects the domain representations, incorporates the incoming E2E delay SLA as contextual input, and outputs predicted decomposition ratios $\hat{\boldsymbol{\phi}^t} = \{\hat{\phi}^t_n\}_{n=1}^{N}$ as described in~\eqref{eq:ratio}.
    \item [4.] In parallel, RADE computes the target decomposition $\{ \tau_n^{*,t} \}_{n=1}^{N}$ for the same SLA request using its risk models and optimization procedure described in Alg.~\ref{alg:rade}.
    \item [5.] The target decompositions are then normalized into ratios over domains:
    \begin{equation}
        \boldsymbol{\phi}^{*,t} = \text{normSum}(\{ {\tau}_n^{*,t} \}_{n=1}^{N}) = \frac{\{ {\tau}_n^{*,t} \}_{n=1}^{N}}{\sum_{n=1}^{N} {\tau}_n^{*,t}}.
    \end{equation}
    \item [6.] Casformer is trained to minimize the Kullback–Leibler (KL) divergence between the target decomposition ratios $\boldsymbol{\phi}^{*,t}$ and the predicted decomposition ratios $\hat{\boldsymbol{\phi}^t}$:
    \begin{equation}
        D_{\mathrm{KL}}\left( \boldsymbol{\phi}^{*,t} \,\middle\|\, \hat{\boldsymbol{\phi}^t} \right) = \sum_{n=1}^{N} \phi_n^{*,t} \log \frac{\phi_n^{*,t}}{\hat{\phi_n^t}}.
    \end{equation}
    Although $\hat{\boldsymbol{\phi}^t}$ and $\boldsymbol{\phi}^{*,t}$ represent decomposition ratios rather than probability distributions \textit{per se}, they are non-negative and sum to one. This structural similarity allows them to be interpreted as categorical distributions over domains. As a result, we can meaningfully apply the KL divergence to measure the distance between the predicted and target decomposition ratios.
\end{enumerate}
The complete high-level training procedure is presented in Alg.~\ref{alg:Casformer-training}. Through this process, Casformer implicitly learns the underlying decomposition policy encoded by RADE, while incorporating raw domain feedback into its single-step predictive model.
This design follows the DINN paradigm, where expert-driven logic (i.e., RADE's optimization constraint and results) guides the training of a data-driven model. Furthermore, since training does not need to be conducted online, training samples can also be collected and reused for offline training.

\begin{algorithm}[h]
\caption{Training Procedure}
\label{alg:Casformer-training}
\begin{algorithmic}[1]
\REQUIRE Casformer $\mathcal{F}$; RADE $\mathcal{R}$; Memory buffers $\{\mathcal{B}_n \}_{n=1}^N$
\FOR{each time step $t = 1$ to $T$}
    \STATE Collect recent feedback and update $\{ \mathcal{B}_n^t \}_{n=1}^N$
    \STATE Receive new E2E SLA request $\tau_{\text{e2e}}^t$
    \STATE $\hat{\boldsymbol{\phi}}^t \gets \mathcal{F}(\{ \mathcal{B}_n^t \}_{n=1}^N, \tau_{\text{e2e}}^t)$ \hfill // Casformer prediction
    \STATE $\boldsymbol{\phi}^{*,t} \gets \text{normSum}(\mathcal{R}(\tau_{\text{e2e}}^t))$ \hfill // RADE computes optima
    \STATE Compute loss: $\mathcal{L}_{\text{train}} = D_{\mathrm{KL}}(\boldsymbol{\phi}^{*,t} \,\|\, \hat{\boldsymbol{\phi}}^t)$
    \STATE Update $\mathcal{F}$ by minimizing $\mathcal{L}_{\text{train}}$
\ENDFOR
\end{algorithmic}
\end{algorithm}

\vspace{1mm}
\noindent \textbf{Inference Phase.} Once trained, Casformer operates independently of RADE. Specifically, at each time step:
\begin{enumerate}
    \item [1.]  The feedback sequences from each domain’s buffer are fed into their respective domain-specific encoders.
    \item [2.]  The domain-specific encoders produce fixed-length feedback representations as described in~\eqref{eq:enc}, which are passed to the cross-domain aggregator.
    \item [3.]  The cross-domain aggregator directly outputs the predicted decomposition assignment in a single forward pass, as depicted in~\eqref{eq:delay_embedding}-\eqref{eq:scale}.
\end{enumerate}
The complete high-level inference procedure is summarized in Alg.~\ref{alg:Casformer-inference}. This forward-only inference pipeline eliminates the need for runtime optimization, risk model maintenance, and gradient updates. As a result, Casformer enables significantly faster decision-making while preserving the decomposition quality learned from RADE.
\begin{algorithm}[h]
\caption{Inference Procedure}
\label{alg:Casformer-inference}
\begin{algorithmic}[1]
\REQUIRE Casformer $\mathcal{F}$; Memory buffers $\{ \mathcal{B}_n \}_{n=1}^N$
\FOR{each time step $t = 1$ to $T$}
    \STATE Collect recent feedback and update $\{ \mathcal{B}_n^t \}_{n=1}^N$
    \STATE Receive new E2E SLA request $\tau_{\text{e2e}}^t$
    \STATE $\hat{\boldsymbol{\phi}}^t \gets \mathcal{F}(\{ \mathcal{B}_n^t \}_{n=1}^N, \tau_{\text{e2e}}^t)$ \hfill // Predict ratios
    \STATE Generate SLA assignments: $\{ \hat{\tau}_n^t \}_{n=1}^N \gets \hat{\boldsymbol{\phi}}^t \cdot \tau_{\text{e2e}}^t$

    \STATE Return $\{ \hat{\tau}_n^t \}_{n=1}^N$
\ENDFOR
\end{algorithmic}
\end{algorithm}

\vspace{1mm}
\noindent \textbf{An Amortized Optimization Perspective.} Functionally, Casformer serves as a \emph{amortization model}~\cite{amortized}, which directly maps E2E SLA requests and domain feedback to decomposition decisions. Particularly, this mapping can be learned by minimizing the expected KL divergence between the predicted decomposition ratios and those produced by RADE:
\begin{equation}
\mathcal{F}_\theta^* = \arg\min_\theta \, \mathbb{E}_{(\{ \mathcal{B}_n \}, \tau_{\text{e2e}})} \left[ D_{\mathrm{KL}} \left( \mathcal{F}_\theta(\{\mathcal{B}_n\}, \tau_{\text{e2e}}), \boldsymbol{\phi}^* \right) \right].
\end{equation}
This approach effectively amortizes the cost of optimization-based decomposition over time with a neural approximation, enabling fast, forward-only inference at deployment while preserving the structural logic of RADE’s optimization behavior.

In summary, Casformer offers a principled, single-stage prediction framework for SLA decomposition that integrates domain knowledge, feedback-driven modeling, and architectural constraints into a cascaded Transformer design. 

\section{Experimental Setup}\label{sec:perf_eval}

\subsection{Evaluation Scenarios}
To evaluate our SLA decomposition framework under realistic conditions, we consider the time-varying simulation environment proposed in~\cite{rails}, where the admission control of each domain for accepting SLA requests is driven by its capacity and system load.
We conduct simulations over \( T = 200 \) discrete time steps using the traffic-driven environment described in Section~\ref{sec:perf_eval}. At each time step, a new E2E service request arrives with a delay budget uniformly sampled from the interval \([90, 110]\)\,ms. 
Particularly, at each time step \( t \), Casformer predicts a decomposition assignment \( \{ d_i^t \}_{i=1}^N \), which is then evaluated using the ground-truth domain acceptance models to compute the actual E2E acceptance probability. The overall performance is then measured by averaging the E2E acceptance probability across all time steps, as defined in~\eqref{eq:decompose-time}.
Casformer is first trained on a \emph{training trace}, during which it runs in parallel with RADE. In this phase, RADE continuously updates risk models and computes optimal decompositions, while Casformer learns to approximate the outputs of RADE from raw domain-specific feedback. After training convergence, we evaluate Casformer independently on a \emph{testing trace}, where it operates without RADE.
All baseline methods are also evaluated on the same testing trace for fair comparison. We compare Casformer against the following baselines:

\vspace{1mm}
\noindent \textbf{Non-Risk-Aware (NRA).} A naive baseline where the E2E delay budget is evenly decomposed across domains. No learning or feedback is utilized. This baseline serves to highlight the importance of risk modeling and adaptive decomposition by providing a simple, uninformed reference point.

\vspace{1mm}
\noindent \textbf{RADE.} A state-of-the-art decomposition framework~\cite{hsu2024online} that uses continuously updated risk models and online optimization, including heuristic search and SLSQP, to compute the decomposition at each time step, as described in Section~\ref{sec:method-bg}.

\vspace{1mm}
\noindent \textbf{Optimal (OPT).} The theoretical upper-bound approach that uses exhaustive search with full knowledge of ground-truth acceptance functions to compute the best possible decomposition at each step.


\subsection{Configurations}
We simulate a network slicing scenario involving \( N = 3 \) domains to evaluate the performance of the proposed Casformer framework and baseline methods. To reflect heterogeneous and dynamic network conditions, each domain is parameterized using randomly sampled values within realistic ranges, following the configurations detailed in~\cite{rails}.
For the RADE framework, configurations adhere to those specified in~\cite{hsu2024online}. For the decomposition process, a heuristic search is first performed by uniformly sampling $10,000$ candidate solutions from the feasible space. The best candidate is then used as the initial solution for the SLSQP algorithm for further refinement.
For Casformer, both the domain-specific feedback encoders (Layer $1$) and the cross-domain aggregator (Layer $2$) are implemented as a $2$-layer Transformer encoder with hidden size $16$ and MLP size $64$. Training uses the AdamW optimizer with the same learning rate of \( 0.01 \), and each domain maintains a memory buffer of size $100$.
All experiments are repeated over $100$ independent simulation runs with different random seeds, and reported results are averaged to ensure statistical significance. 
The simulation environment was implemented in Python 3.11.9, with models built using PyTorch 2.5.1. All experiments ran on a server equipped with an Intel Core i7-10700K CPU, 32 GB RAM, and an RTX 2080S GPU.

\section{Results and Discussion}\label{sec:results}
\subsection{Long-Term Performance and Efficiency}
To evaluate both the effectiveness and efficiency of Casformer, we compare its performance against several baselines in terms of two key metrics: ($i$) the long-term average E2E SLA acceptance probability, denoted $\overline{p}_{\text{e2e}}$ as defined in~\eqref{eq:decompose-time}, and ($ii$) the average runtime required to compute a decomposition decision at inference time.
\begin{figure}[h]
    \centering
    \begin{subfigure}{0.32\textwidth}
        \centering
        \includegraphics[width=\textwidth, trim=0.2cm 0cm 0cm 0.2cm, clip]{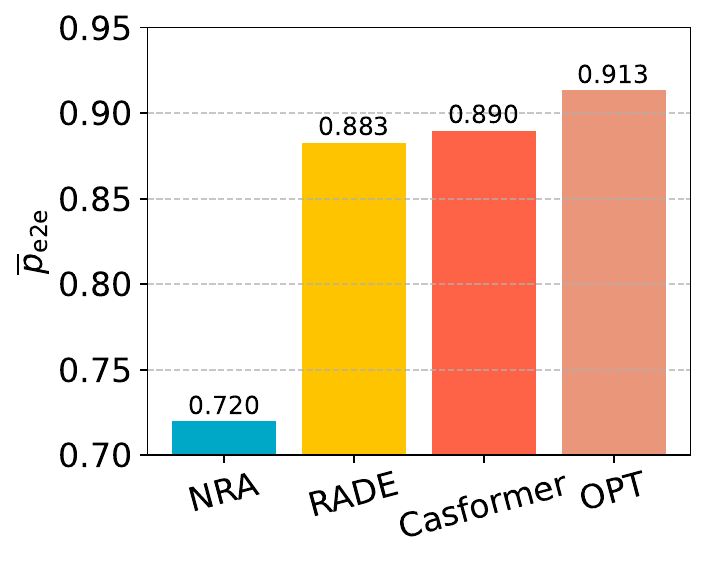} 
        \caption{E2E Acceptance Probability}
        \label{fig:e2eacc}
    \end{subfigure}
    \hspace*{-0.2cm}
    \begin{subfigure}{0.164\textwidth}
        \centering
        \includegraphics[width=\textwidth, trim=0.2cm 0.2cm 0cm 0cm, clip]{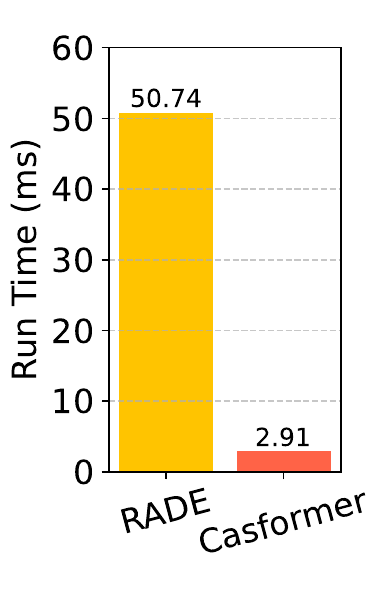} 
        \caption{Run Time}
        \label{fig:runtime}
    \end{subfigure}
    \caption{Performance comparison: (a) Average E2E SLA acceptance probability over time; (b) run time per decomposition.}
    \label{fig:combined}
\end{figure}
Fig.~\ref{fig:e2eacc} presents the average E2E acceptance probability over the entire simulation horizon. 
Both Casformer and RADE approach the optimal upper bound. Casformer achieves an average acceptance probability of $0.89$, slightly outperforming its teacher model RADE, and significantly surpassing the simple baseline NRA. 
While Casformer is trained to imitate RADE's decisions, it unexpectedly exceeds the performance of RADE. 
This is interesting, as student models are generally designed to match, not exceed, the performance of their teacher.
We argue that this improvement stems from Casformer's ability to generalize well to unseen conditions, which smooths over noisy or overfitting behavior in RADE. 
Further investigation into generalizability is presented in Section~\ref{sec:general}.

Fig.~\ref{fig:runtime} shows the average runtime per decomposition decision. Casformer requires only $2.91$\,ms per inference, which is over $17$ times faster than RADE. This significant reduction in computational cost is achieved by eliminating the need for online optimization, and instead performing a single forward pass through the learned model for decomposition.
In summary, these results validate that Casformer achieves competitive or even superior long-term decision quality compared to RADE while enabling fast SLA decomposition at runtime.
\subsection{Model Generalizability}\label{sec:general}
\begin{figure}[h]
    \centering
    \begin{subfigure}{0.23\textwidth}
        \centering
        \includegraphics[width=\textwidth, trim=0.2cm 0cm 0.2cm 0.2cm, clip]{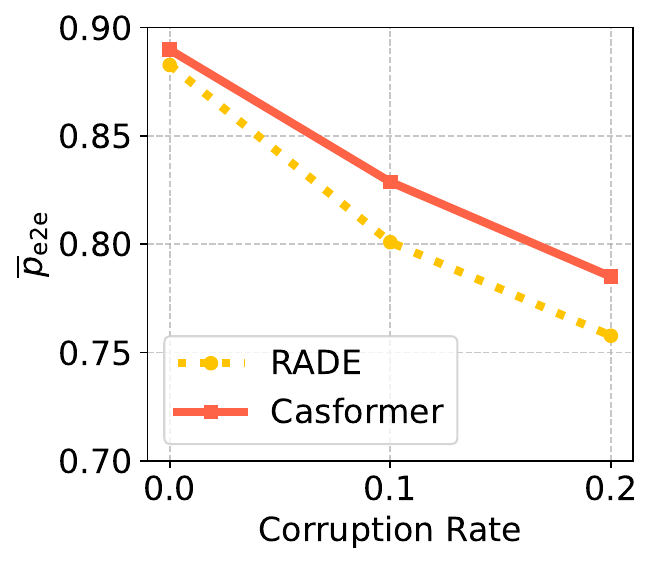} 
        \caption{Data Corruption}
        \label{fig:corrupt}
    \end{subfigure}
    \hspace*{-0.1cm}
    \begin{subfigure}{0.23\textwidth}
        \centering
        \includegraphics[width=\textwidth, trim=0.2cm 0cm 0.2cm 0.2cm, clip]{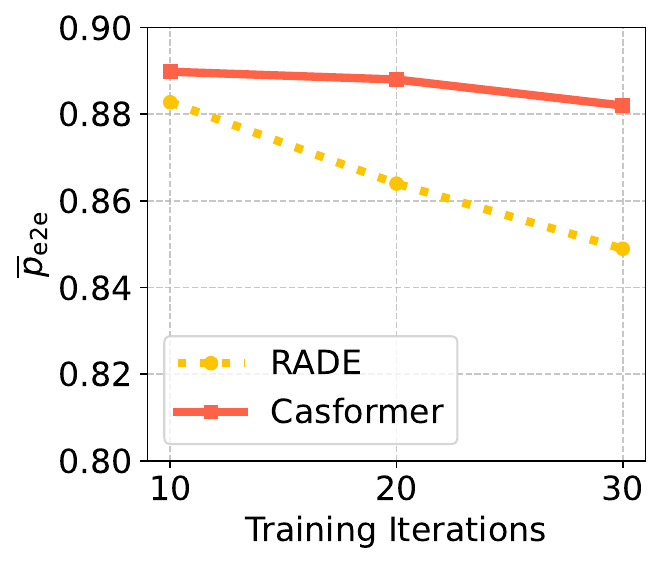} 
        \caption{Target Overfitting}
        \label{fig:overfit}
    \end{subfigure}
    \caption{Generalizability comparison under two conditions: (a) input-level corruption; (b) target-level overfitting.}
    \label{fig:combined}
\end{figure}

To examine generalization ability under imperfect training signals, we consider two complementary stress tests targeting different aspects of generalization: ($i$) input-level corruption and ($ii$) target-level overfitting.
Fig.~\ref{fig:corrupt} presents the results under increasing levels of feedback corruption, where each historical sample has a fixed probability of label flipping. 
This setting simulates temporary anomalies in domain admission control due to transient network conditions such as misconfigurations, system faults, or outages.
As the corruption rate increases, the performance of both models declines. However, Casformer consistently maintains a higher average acceptance probability than RADE, indicating greater robustness to input-level noise. We attribute this to the regularization effect of amortized optimization~\cite{amortized}, which encourages smoother representations and mitigates the influence of outliers.
Fig.~\ref{fig:overfit} explores the sensitivity to overfitting. Specifically, we vary the number of OGD iterations for online risk model update in RADE. While additional updates help the model fit better to the buffered data, they also increase the risk of overfitting to the limited contents of the memory buffer. As shown, the performance of RADE degrades as training iterations increase, while Casformer remains largely unaffected, highlighting its ability to capture useful general patterns beyond the specific training trajectory.
Together, these results demonstrate that Casformer achieves improved generalizability over RADE across both noisy input conditions and target overfitting risks. This robustness underpins its effectiveness in unseen scenarios.

\subsection{Computational Scalability}
\begin{figure}[h]
    \centering
    \begin{subfigure}{0.21\textwidth}
        \centering
        \includegraphics[width=\textwidth, trim=0.2cm 0cm 0.2cm 0.2cm, clip]{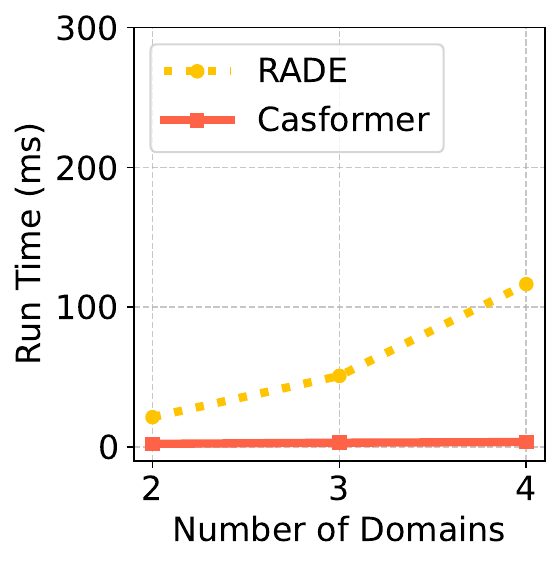} 
        \caption{GPU}
        \label{fig:scale_gpu}
    \end{subfigure}
    \hspace*{0.3cm}
    \begin{subfigure}{0.21\textwidth}
        \centering
        \includegraphics[width=\textwidth, trim=0.2cm 0cm 0.2cm 0.2cm, clip]{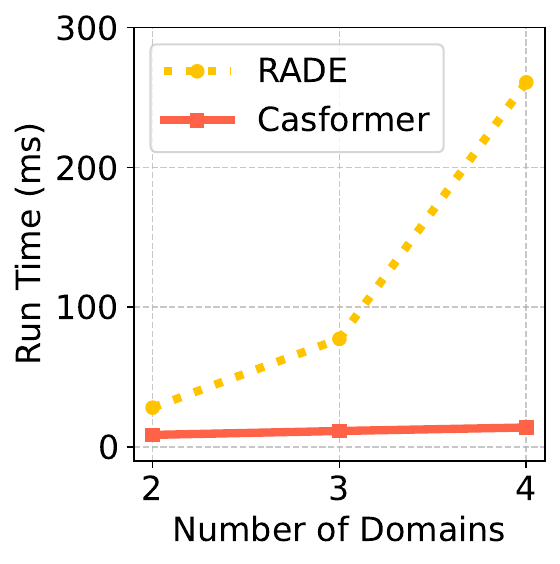} 
        \caption{CPU}
        \label{fig:scale_cpu}
    \end{subfigure}
    \caption{Inference scalability with increasing number of domains: Average runtime on (a) GPU and (b) CPU.}
    \label{fig:scalability}
\end{figure}

Fig.~\ref{fig:scalability} shows the inference runtime on both GPU and CPU for scenarios involving $2$, $3$, and $4$ domains.
As seen in the figure, Casformer maintains near-constant inference time as the number of domains increases, thanks to its forward-only architecture and parallelized Transformer-based design. In contrast, RADE’s runtime increases significantly with more domains, especially on CPU, as RADE must solve increasingly complex optimization problems at runtime by design.
These results indicate that Casformer is far more scalable than RADE in inference speed, making it more suitable for real-time applications in multi-domain network systems.

\section{Conclusion}\label{sec:conclusion}
In this paper, we addressed a critical challenge of decomposing E2E SLAs for network slicing across multiple domains. 
Existing SLA decomposition approaches rely heavily on iterative optimization and frequent online updates of domain-specific risk models, imposing substantial computational overhead and latency. 
To overcome these limitations, we introduced Casformer, a Risk-Informed Cascaded Transformer that supports optimization-free SLA decomposition.
Casformer innovatively leverages a cascaded Transformer architecture and a learning paradigm inspired by DINNs, enabling direct prediction of SLA decomposition decisions through a single forward pass.
This approach significantly reduces system complexity by eliminating the need for online optimization and updating of risk models.
Extensive experiments demonstrate that Casformer achieves strong scalability and generalizability while maintaining high-quality SLA decomposition under noisy and volatile network environments. In addition, its forward-only inference reduces runtime complexity against optimization-based baselines, simplifying deployment without sacrificing decision quality.
Overall, our work highlights the potential of amortized optimization and Transformer architectures for network automation, providing a scalable, efficient, and practically deployable solution for SLA management in emerging 5G-and-beyond networks.

\section*{Acknowledgment}
This research was partially funded by the Dutch 6G flagship project ``Future Network Services''.


\bibliographystyle{IEEEtran}
\bibliography{references.bib}

%



\end{document}